\documentclass[aps,prc,showpacs,twocolumn,amssymb,superscriptaddress,floatfix]{revtex4}

\usepackage[colorlinks,citecolor=blue]{hyperref}
\usepackage{amsmath, amsthm, amssymb}
\usepackage{graphicx}

\def\be{\begin{eqnarray}}
\def\ee{\end{eqnarray}}
\def\lsim{\stackrel{\scriptstyle <}{\phantom{}_{\sim}}}
\def\gsim{\stackrel{\scriptstyle >}{\phantom{}_{\sim}}}

\begin{document}

\title{Azimuthal anisotropies\\
for Au+Au collisions in the parton-hadron transient energy range}

\author{V. P. Konchakovski}
\affiliation{Institute for Theoretical Physics, University of Giessen, Giessen, Germany}
\affiliation{Bogolyubov Institute for Theoretical Physics, Kiev, Ukraine}

\author{E. L. Bratkovskaya}
\affiliation{Institute for Theoretical Physics, University of Frankfurt, Frankfurt, Germany}
\affiliation{Frankfurt Institute for Advanced Studies, Frankfurt, Germany}

\author{W. Cassing}
\affiliation{Institute for Theoretical Physics, University of Giessen, Giessen, Germany}

\author{V. D. Toneev}
\affiliation{Frankfurt Institute for Advanced Studies, Frankfurt, Germany}
\affiliation{Joint Institute for Nuclear Research, Dubna, Russia}

\author{S. A. Voloshin}
\affiliation{Wayne State University, Detroit, Michigan, USA}

\author{V. Voronyuk}
\affiliation{Bogolyubov Institute for Theoretical Physics, Kiev, Ukraine}
\affiliation{Frankfurt Institute for Advanced Studies, Frankfurt, Germany}
\affiliation{Joint Institute for Nuclear Research, Dubna, Russia}

\begin{abstract}
The azimuthal anisotropies of the collective transverse flow of
charged hadrons are investigated in a wide range of heavy-ion
collision energies within the microscopic parton-hadron-string
dynamics (PHSD) transport approach, which incorporates explicit
partonic degrees of freedom in terms of strongly interacting
quasiparticles (quarks and gluons) in line with an equation of state
from lattice QCD as well as the dynamical hadronization and hadronic
collision dynamics in the final reaction phase. The experimentally
observed increase of the elliptic flow $v_2$ of charged hadrons with
collision energy is successfully described in terms of the PHSD
approach. The PHSD scaling properties of various collective
observables are confronted with experimental data as well as with
hydrodynamic predictions. The analysis of higher-order harmonics $v_3$
and $v_4$ in the azimuthal angular distribution shows a similar
tendency of growing deviations between partonic and purely hadronic
models with increasing collision energy. This demonstrates that the
excitation functions of azimuthal anisotropies reflect the increasing
role of quark-gluon degrees of freedom in the early phase of
relativistic heavy-ion collisions. Furthermore, the specific variation
of the ratio $v_4/(v_2)^2$ with respect to bombarding energy,
centrality, and transverse momentum is found to provide valuable
information on the underlying dynamics.
\end{abstract}


\maketitle

\section{Introduction}

The discovery of large azimuthal anisotropic flow at the Relativistic
Heavy Ion Collider (RHIC) provides conclusive evidence for the
creation of dense partonic matter in ultra relativistic
nucleus-nucleus collisions. With sufficiently strong parton
interactions, the medium in the collision zone can be expected to
achieve local equili\-brium and exhibit approximately hydrodynamic
flow~\cite{Ol92,HK02,Sh09}. The momentum anisotropy is generated due
to pressure gradients of the initial ``almond-shaped'' collision zone
produced in noncentral collisions~\cite{Ol92,HK02}.  The azimuthal
pressure gradient extinguishes itself soon after the start of the
hydrodynamic evolution, so the final flow is insensitive to later
stages of the fireball evolution. The pressure gradients have to be
large enough to translate an early asymmetry in density of the initial
state to a final-state momentum-space anisotropy. In these collisions
a new state of strongly interacting matter is created, being
characterized by a very low shear viscosity $\eta$ to entropy density
$s$ ratio, $\eta/s$, close to a nearly perfect
fluid~\cite{Sh05,GMcL05,PC05}. Lattice QCD (lQCD)
calculations~\cite{zolt,Cheng08,aori10} indicate that a crossover
region between hadron and quark-gluon matter should have been reached
in these experiments.

An experimental manifestation of this collective flow is the
anisotropic emission of charged particles in the plane transverse to
the beam direction. This anisotropy is described by the different flow
parameters defined as the proper Fourier coefficients $v_n$ of the
particle distributions in azimuthal angle $\psi$ with respect to the
reaction plane angle $\Psi_{RP}$. At the highest RHIC collision energy
of $\sqrt{s_{NN}} =$ 200~GeV, differential elliptic flow measurements
$v_2(p_T)$ have been reported for a broad range of centralities or
number of participants $N_{part}$. For $N_{part}$ estimates, the
geometric fluctuations associated with the positions of the nucleons
in the collision zone serve as the underlying origin of the initial
eccentricity fluctuations. These data are found to be in accord with
model calculations that an essentially locally equilibrated
quark-gluon plasma (QGP) has little or no
viscosity~\cite{HK02,PHEN07,RR07,XGS08}. Collective flow continues to
play a central role in characterizing the transport properties of the
strongly interacting matter produced in heavy-ion collisions at
RHIC. Particle anisotropy measurements are considered as key
observables for a reliable extraction of transport coefficients.

A quark-number scaling of the elliptic-flow data is observed for a
broad range of particle species, collision centralities, and
transverse kinetic energy, which is interpreted as being due to the
development of substantial collectivity in the partonic
phase~\cite{Lac07}. Small violations of the scaling of $v_2(N_{part})$
with the initial eccentricity of the collision zone $\epsilon_2$
suggest a strongly coupled low-viscosity plasma ${\eta}/{s}\sim
(1-2)/(4\pi)$ in energetic Au+Au
collisions~\cite{Lac07,DDGO07,Ch10}. The initial eccentricity of the
collision zone (and its associated fluctuations) has proven to be an
essential ingredient for these extractions. Nevertheless, the degree
to which the QGP is thermalized is still being debated~\cite{Bo06}.

It was shown before that higher-order anisotropy harmonics, in
particular $v_4$, may provide a more sensitive constraint on the
magnitude of $\eta/s$ and the freeze-out dynamics, and the ratio
$v_4/(v_2)^2$ might indicate whether a full local equilibrium is
achieved in the QGP~\cite{Bh05}. The role of fluctuations and
so-called nonflow correlations are important for such measurements. It
is well established that initial eccentricity fluctuations
significantly influence the magnitudes of
$v_{2,4}$~\cite{MS03}. However, the precise role of nonflow
correlations, which lead to a systematic error in the determination of
$v_{2,4}$, is less clear. Recently, significant attention has been
given to the study of the influence of initial geometry fluctuations
on higher order eccentricities $\epsilon_n (n\geq 3$) for a better
understanding of how such fluctuations manifest themselves in the
harmonic flow correlations characterized by $v_n$.  Even more, it was
proposed that the analysis of $v_n^2$ for all values of $n$ can be
considered as an analogous measurement to the power spectrum extracted
from the cosmic microwave background Radiation providing a possibility
to observe superhorizon fluctuations~\cite{MMSS08}.

More recently, the importance of the triangular flow $v_3$, which
originates from fluctuations in the initial collision geometry, has
been pointed out~\cite{AR10,XK10,Pet123}. The participant
triangularity characterizes the triangular anisotropy of the initial
nuclear overlap geometry. It arises from event-by-event fluctuations
in the participant-nucleon collision space-time points and
corresponds to a large third Fourier component in the two-particle
azimuthal correlations at large pseudo rapidity separation $\Delta
\eta$. This fact suggests a significant contribution of the
triangular flow to the ridge phenomenon and broad away-side
structures observed in the RHIC data~\cite{jets}. The ridge might be
related to flux-tube-like structures in the initial state as argued
in Ref.~\cite{Vol05} or successive coherent gluon radiation as
suggested in Ref.~\cite{Carsten}.

A large number of anisotropic flow measurements have been performed by
many experimental groups at SIS (Schwerionensynchroton), AGS
(Alternating Gradient Synchrotron), SPS (Super Proton Synchrotron),
and RHIC (Relativistic Heavy Ion Collider) energies over the last
twenty years. Very recently, the azimuthal asymmetry has been measured
also at the LHC (Large Hadron Collider)~\cite{LHC}. However, the fact
that these data have not been obtained under the same experimental
conditions as at RHIC experiments, does not directly allow for a
detailed and meaningful comparison in most cases. The experimental
differences include different centrality selection, different
transverse momentum acceptance, different particle species, different
rapidity coverage, and different methods for flow analysis as pointed
out in Ref.~\cite{Tar11}.

The Beam Energy Scan (BES) program proposed at RHIC~\cite{ST11}
covers the energy interval from $\sqrt{s_{NN}}=$ 200~GeV, where
partonic degrees of freedom play a decisive role, down to the AGS
energy of $\sqrt{s_{NN}}\approx$ 5~GeV, where most experimental data
may be described successfully in terms of hadronic
degrees of freedom, only. Lowering the RHIC collision energy and
studying the energy dependence of anisotropic flow allows us to search
for the possible onset of the transition to a phase with partonic
degrees of freedom at an early stage of the collision, as well as
possibly to identify the location of the critical end point that
terminates the crossover transition at small quark-chemical
potential to a first-order phase transition at higher quark-chemical
potential~\cite{Lac07,Agg07}.

This work aims to study excitation functions for different harmonics
of the charged-particle anisotropy in the azimuthal angle at
midrapidity in a wide transient energy range, {\it i.e.}, from the AGS
to the top RHIC energy. The first attempts to explain the preliminary
STAR data with respect to the observed increase of the elliptic flow
$v_2$ with the collision energy have failed, since the traditional
available models did not allow clarification of the role of the
partonic phase~\cite{NKKNM10}. In this paper, as an extension of our
recent study in Ref.~\cite{v2short}, we investigate the energy behavior
of different flow coefficients, their scaling properties and
differential distributions. Our analysis of the STAR/PHENIX RHIC data
-- based on recent results of the BES program -- will be performed
within the parton-hadron-string dynamics (PHSD) transport
model~\cite{PHSD} that includes explicit partonic degrees of freedom
as well as a dynamical hadronization scheme for the transition from
partonic to hadronic degrees of freedom and vice versa.

The paper is organized as follows: In Section~\ref{sec2} we will briefly
recall the main ingredients of the PHSD approach as well as the
performance of PHSD for relativistic heavy-ion collisions from the
lower SPS to the top RHIC energies. Section~\ref{sec3} is devoted to the
actual results from PHSD for the excitation function of the elliptic
flow $v_2$ in comparison to the hadron-string dynamics (HSD)
approach and other related models, as well as to the available data
from the STAR and PHENIX Collaborations. We also provide results
from the PHSD and HSD models for the excitation functions of $v_3$
and $v_4$ in view of the Beam Energy Scan (BES) program at RHIC in
order to identify partonic contributions. Scaling properties of
experimental data, in particular the universal and longitudinal
scaling relations found empirically, are elaborated here and
compared to a hydrodynamic description. To be more specific, we will
also present the calculated results for the $p_T$ dependence of
elliptic flow at midrapidity for minimum bias collisions of Au+Au
for $\sqrt{s_{NN}}$ from 5 to 200~GeV. Furthermore, the centrality
dependence of $v_2$, $v_3$, and $v_4$ will be addressed at the top RHIC
energy. Section~\ref{sec4} provides the conclusions of our present study and
indicates the open problems.

\section{The PHSD model}
\label{sec2}

\begin{figure}[thb]
\includegraphics[width=\linewidth]{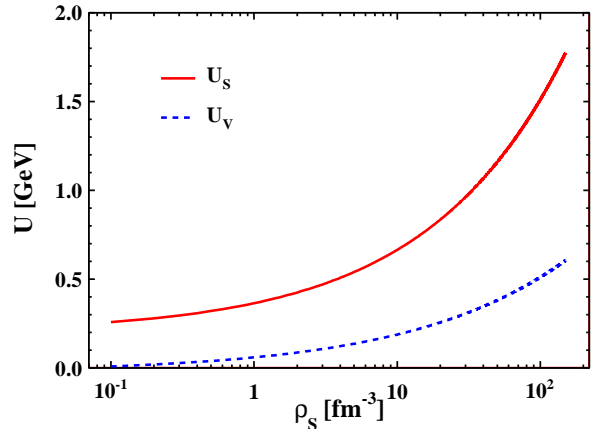}
\caption{(Color online) The scalar and vector mean-field potentials in
  the present PHSD model as a function of the scalar density $\rho_s$
  of partons.}
\label{meanF}
\end{figure}

The dynamics of partons, hadrons and strings in relativistic
nucleus-nucleus collisions is analyzed within the novel
parton-hadron-string dynamics (PHSD) approach~\cite{PHSD,BCKL11}. In
this transport approach the partonic dynamics is based on the
dynamical quasiparticle model (DQPM)~\cite{Cassing06,Cassing07},
which describes QCD properties in terms of single-particle Green's
functions [in the sense of a two-particle irreducible (2PI)
approach]. In Ref.~\cite{BCKL11} the actual (essentially three) DQPM
parameters for the temperature-dependent effective coupling have
been fitted to the recent lattice QCD results of Ref.~\cite{aori10}.
The latter lead to a critical temperature $T_c \approx$ 160~MeV,
which corresponds to a critical energy density of $\varepsilon_c
\approx$ 0.5~GeV/fm$^3$. In PHSD the parton spectral functions
$\rho_j$ ($j=q, {\bar q}, g$) are no longer $\delta$-functions in
the invariant mass squared (as in conventional cascade or transport
models), but are taken as
\begin{equation}
 \rho_j(\omega,{\bf p})
 =
 \frac{\gamma_j}{E_j} \left(
   \frac{1}{(\omega-E_j)^2+\gamma_j^2} - \frac{1}{(\omega+E_j)^2+\gamma_j^2}
 \right)
\label{eq:rho}
\end{equation}
separately for quarks/antiquarks and gluons ($j=q,\bar{q},g$).
With the convention $E^2({\bf p}^2) = {\bf p}^2+M_j^2-\gamma_j^2$, the
parameters $M_j^2$ and $\gamma_j$ are directly related to the real
and imaginary parts of the retarded self-energy, {\it e.g.}, $\Pi_j =
M_j^2-2i\gamma_j\omega$. The spectral function~(\ref{eq:rho}) is
antisymmetric in $\omega$ and is normalized as
\begin{equation}
 \int_{-\infty}^{\infty} \frac{d \omega}{2 \pi} \
 \omega \ \rho_j(\omega, {\bf p}) = \int_0^{\infty} \frac{d
 \omega}{2 \pi} \ 2 \omega \ \rho_j(\omega, {\bf p}) = 1 \ .
\label{normalize}
\end{equation}
The actual parameters in Eq.~(\ref{eq:rho}), {\it i.e.}, the gluon
mass $M_g$ and width $\gamma_g$ -- employed as input in the PHSD
calculations -- as well as the quark mass $M_q$ and width $\gamma_q$,
are depicted in Fig.~\ref{meanF} of Ref.~\cite{BCKL11} as a function
of the scaled temperature $T/T_c$. As mentioned above, these values
for the masses and widths have been fixed by fitting the lattice QCD
results from Ref.~\cite{aori10} in thermodynamic equilibrium. We
recall that the DQPM allows extraction of a potential energy density
$V_p$ from the space-like part of the energy-momentum tensor, which
can be tabulated, {\it e.g.}, as a function of the scalar parton
density $\rho_s$. Derivatives of $V_p$ with respect to $\rho_s$ then
define a scalar mean-field potential $U_s(\rho_s)$, which enters the
equation of motion for the dynamical partonic quasiparticles. As one
can see from Fig.~\ref{meanF}, the scalar potential is rather large
and nonlinearly increases with $\rho_s$. This implies that the
repulsive force due to $U_s(\rho_s)$ will change in a nonmonotonous
way with the scalar density. The vector mean-field potential is not
negligible too, especially at high $\rho_s$, and induces a Lorentz
force for the partons. Note that the vector mean-field potential
vanishes with decreasing scalar density whereas the scalar mean-field
potential approaches a constant value for $\rho_s\rightarrow 0$.

Furthermore, a two-body interaction strength can be extracted from the
DQPM as well from the quasiparticle width in line with
Ref.~\cite{PC05} ({\it cf.}  Refs.~\cite{PHSD,BCKL11} for
details). The transition from partonic to hadronic degrees of freedom
(and vice versa) is described by covariant transition rates for the
fusion of quark-antiquark pairs or three quarks (antiquarks),
respectively, obeying flavor current-conservation, color neutrality, as
well as energy-momentum conservation. Since the dynamical quarks and
antiquarks become very massive close to the phase transition, the
formed resonant ``pre-hadronic'' color-dipole states ($q\bar{q}$
or $qqq$) are of high invariant mass too, and sequentially decay to
the ground-state meson and baryon octets, increasing the total entropy.

On the hadronic side PHSD includes explicitly the baryon octet and
decouplet, the $0^-$- and $1^-$-meson nonets, as well as selected
higher resonances as in the hadron-string dynamics (HSD)
approach~\cite{Ehehalt,HSD}. Hadrons of higher masses ($>$1.5~GeV in the
case of baryons and $>$1.3~GeV in the case of mesons) are treated as
``strings'' (color dipoles) that decay to the known (low-mass) hadrons
according to the JETSET algorithm~\cite{JETSET}. We discard an
explicit recapitulation of the string formation and decay and refer
the reader to the original work~\cite{JETSET}. Note that PHSD and HSD
(without explicit partonic degrees of freedom) merge at low energy
density, in particular below the critical energy density
$\varepsilon_c\approx$ 0.5~GeV/fm$^{3}$.

The PHSD approach has been applied to nucleus-nucleus collisions
from $\sqrt{s_{NN}}\sim$ 5 to 200~GeV in Refs.~\cite{PHSD,BCKL11} in
order to explore the space-time regions of ``partonic matter''. It was
found that even central collisions at the top SPS energy of
$\sqrt{s_{NN}}=$ 17.3~GeV show a large fraction of nonpartonic, {\it
i.e.}, hadronic or string-like matter, which can be viewed as a
hadronic corona. This finding implies that neither hadronic nor only
partonic ``models'' can be employed to extract physical conclusions in
comparing model results with data. In addition, we have found that
the partonic phase has a low impact on rapidity distributions of
hadrons but a sizable influence on the transverse mass distribution
of final kaons due to the repulsive partonic mean
fields~\cite{PHSD}. It has been, furthermore, demonstrated in
Ref.~\cite{BCKL11} that at $\sqrt{s_{NN}}=$ 200~GeV the PHSD model
gives a reasonable reproduction of hadron rapidity distributions and
transverse mass spectra, and also a fair description of the elliptic
flow of charged hadrons as a function of the centrality of the
reaction and the transverse momentum $p_T$.

Furthermore, an approximate quark-number scaling of the elliptic flow
$v_2$ of identified hadrons is observed in the PHSD results at top
RHIC energies too. As indicated above, PHSD merges to HSD in the
lower (transient) energy regime. Both approaches are well in line with
experimental data in the lower SPS energy regime as shown in
Ref.~\cite{PHSD}. All these previous findings provide promising
perspectives to use PHSD in the whole range from about
$\sqrt{s_{NN}}=$ 5 to 200~GeV.

\section{Results for collective flows}
\label{sec3}

\subsection{Elliptic flow}

The largest component, known as elliptic flow $v_2$, is one of the
early observations at RHIC~\cite{Ac01}. More recently, it was
noticed that  fluctuations in the initial geometry are very
important~\cite{AR10}.  The elliptic flow coefficient is a widely
used quantity characterizing the azimuthal anisotropy of emitted
particles,
 \be \label{eqv2}
 v_2 = <cos(2\psi-2\Psi_{RP})>=<\frac{p^2_x - p^2_y}{p^2_x + p^2_y}>~,
\ee
where $\Psi_{RP}$ is the azimuth of the reaction plane, $p_x$ and
$p_y$ are the $x$ and $y$ component of the particle momenta, and the
brackets denote averaging over particles and events. This coefficient
can be considered as a function of centrality, pseudorapidity $\eta$,
and/or transverse momentum $p_T$. We note that the reaction plane in
PHSD is given by the $x-z$ plane with the $z$ axis in the beam
direction. The reaction plane is defined as a plane containing the
beam axes and the impact parameter vector.

\begin{figure}[t]
\includegraphics[width=\linewidth]{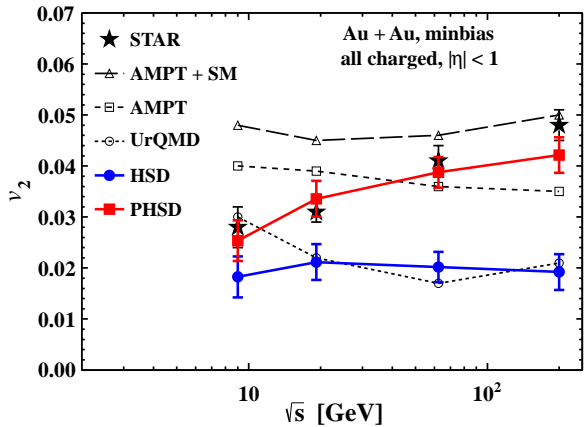}
\caption{(Color online) The average elliptic flow $v_2$ of charged particles at
  midrapidity for minimum bias collisions at $\sqrt{s_{NN}}=$ 9.2,
  19.6, 62.4 and 200~GeV (given by stars) is taken from the data
  compilation of Ref.~\cite{NKKNM10}. The corresponding results from
  different models are compared to the data and explained in more
  detail in the text.}
\label{s}
\end{figure}

We recall that at high bombarding energies the longitudinal size of
the Lorentz contracted nuclei becomes negligible compared to its
transverse size. The forward shadowing effect then goes away and the
elliptic flow fully develops in-plane, leading to a positive value of
the average flow $v_2$ since no shadowing from spectators takes
place. In Fig.~\ref{s} the experimental $v_2$ data compilation for the
transient energy range is compared to the results from HSD
calculations and further available model results as included in
Ref.~\cite{NKKNM10}. The centrality selection is the same for the data
and the various models.

In order to interpret the results in Fig.~\ref{s} we have to recall
the various ingredients of the models employed for comparison. The
UrQMD (ultrarelativistic quantum molecular dynamics) model is a
microscopic transport theory based on the relativistic Boltzmann
equation~\cite{UrQMD}. It allows for the on-shell propagation of all
hadrons along classical trajectories in combination with stochastic
binary scattering, color string formation, and resonance decay. The
model incorporates baryon-baryon, meson-baryon, and meson-meson
interactions based on experimental data (when possible). This
Boltzmann-like hadronic transport model has been employed for
proton-nucleus and nucleus-nucleus collisions from AGS to RHIC
energies~\cite{UrQMD}. The comparison of the data on $v_2$ to those
from the UrQMD model will thus essentially provide information on the
contribution from the hadronic phase. As seen in Fig.~\ref{s}, being
in agreement with data at the lowest energy $\sqrt{s_{NN}}=$ 9.2~GeV,
the UrQMD model results then either remain approximately constant or
decrease slightly with increasing $\sqrt{s_{NN}}$; UrQMD thus does not
reproduce the rise of $v_2$ with the collision energy as seen
experimentally.

The HSD model~\cite{HSD,BCS03} is also a hadron-string model
including formally the same processes as UrQMD. However, being
based on the off-shell generalized transport equation~\cite{Cas09}
followed from Kadanoff-Baym approach, the quasiparticles
in the HSD model take into account in-medium modifications of their
properties in the nuclear environment, which is rather essential for
many observables and in particular for dileptons. Detailed
comparisons between HSD and UrQMD for central Au + Au (Pb + Pb)
collisions have been reported in Refs.~\cite{Weber,BRAT04} from AGS
to top SPS energies with respect to a large experimental data set.
Indeed, both hadronic approaches yield similar results on the level
of 20\%-30\%, which is also the maximum deviation from the data sets.
Accordingly, the HSD model also predicts an approximately
energy-independent flow $v_2$ in quite close agreement with the
UrQMD results. We may thus conclude that the rise of $v_2$ with
bombarding energy is not due to hadronic interactions and models with
partonic degrees of freedom have to be addressed.

The AMPT (a multi phase transport) model~\cite{AMPT} uses initial
conditions of a perturbative QCD (pQCD) inspired model, which produces
multiple minijet partons according to the number of binary initial
nucleon-nucleon collisions. These (massless) minijet partons undergo
scattering (without potentials) before they are allowed to fragment
into hadrons. The string melting (SM) version of the AMPT model
(labeled in Fig.~\ref{s} as AMPT-SM) is based on the idea that the
existence of strings (or hadrons) is impossible for energy densities
beyond a critical value of $\varepsilon\sim$ 1~GeV/fm$^3$. Hence they
need to melt the strings to (massless) partons. This is done by
converting the mesons to a quark and antiquark pair, baryons to three
quarks, {\it etc}, fulfilling energy-momentum conservation. The
subsequent scatterings of the quarks are based on a parton cascade with
(adjustable) effective cross sections that are significantly larger
than those from pQCD~\cite{AMPT}.  Once the partonic interactions
terminate, the partons hadronize through the mechanism of parton
coalescence.

We find from Fig.~\ref{s} that the interactions between the minijet
partons in the AMPT model indeed increase the elliptic flow
significantly as compared to the hadronic models UrQMD and HSD. An
additional inclusion of interactions between partons in the AMPT-SM
model gives rise to another 20\% of $v_2$, bringing it into
agreement (for AMPT-SM) with the data at the maximal collision
energy. So, both versions of the AMPT model indicate the importance
of partonic contributions to the observed elliptic flow $v_2$ but
do not reproduce its growth with $\sqrt{s_{NN}}$. The authors
address this result to the partonic equation of state (EoS) employed,
which corresponds to a massless and noninteracting relativistic gas
of particles. This EoS deviates severely from the results of lattice
QCD calculations for temperatures below 2-3 $T_c$. Accordingly, the
degrees of freedom are propagated without self-energies and a parton
spectral function.

\begin{figure}[thb]
\includegraphics[width=\linewidth]{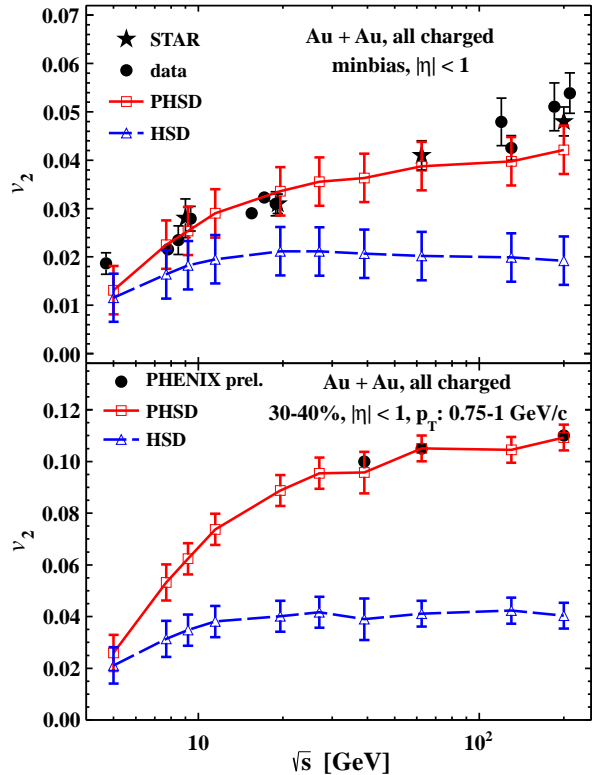}
\caption{(Color online) Average elliptic flow $v_2$ of charged
  particles at mid-pseudorapidity for two centrality selections
  calculated within the PHSD (solid curves) and HSD (dashed lines)
  models. The $v_2$ STAR data (stars) for minimal bias are the same as
  in Fig.~\ref{s} (stars); the preliminary PHENIX
  data~\cite{PHENIX_v2_s} are plotted by filled circles and other data
  are taken from the compilation in Ref.~\cite{AbST10}.}
\label{vns}
\end{figure}

The PHSD approach incorporates the latter medium effects in line
with a lQCD equation-of-state as discussed in Section~\ref{sec2} and also
includes a dynamical hadronization scheme based on covariant
transition rates. As has been shown in our previous study~\cite{v2short},
the elliptic flow $v_2$ from PHSD (red solid lines in Figs.~\ref{s} and
\ref{vns}) agrees with the data from the STAR and PHENIX Collaborations
and clearly shows an increase with bombarding energy. As was
demonstrated in the thorough analysis of Ref.~\cite{Tar11}, the
difference between STAR/PHENIX $v_2$ results is less than 2\%-5\% below
$p_T \simeq$ 2.5~GeV/c. At higher transverse momentum, the STAR
elliptic flow $v_2$ is systematically larger than the PHENIX $v_2$ and
the ratio tends to grow with $p_T$, reaching the value of 20\% at
$p_T\simeq$ 5.5~GeV/c. The differences in $v_2$ at higher $p_T$ might
be attributed to non-flow effects due to di-jets, which are mostly
suppressed by the rapidity gaps in the case of the PHENIX
measurements. Anyhow, we do not consider such high transverse momenta.

Note that PHSD and AMPT-SM practically give the same elliptic flow at
the top RHIC energy of $\sqrt{s_{NN}}=$ 200~GeV. However, PHSD is more
elaborated and includes more realistic properties of dynamical
quasiparticles, especially in the vicinity of the critical energy
density.

\begin{figure}[thb]
\includegraphics[width=\linewidth]{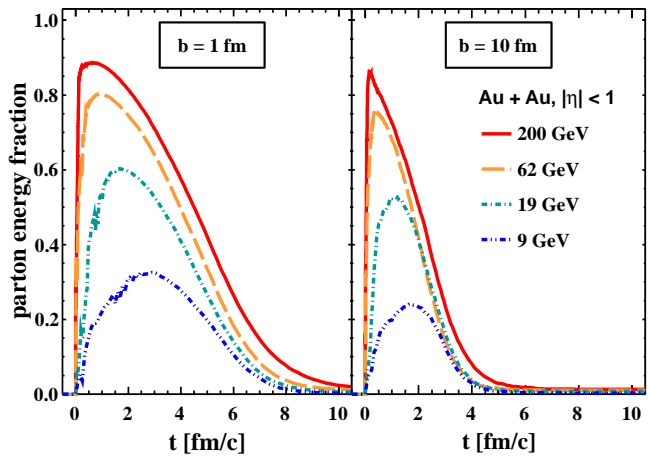}
\caption{(Color online) Evolution of the parton fraction of the total
  energy density at midrapidity for different collision energies at
  impact parameters $b=$ 1 and 10~fm.}
\label{part}
\end{figure}

An explanation for the increase in $v_2$ with collision energy is
provided in Fig.~\ref{part}. Here we show the partonic fraction of the
energy density with respect to the total energy where the energy
densities are calculated at midrapidity. As discussed above, the main
contribution to the elliptic flow is coming from an initial partonic
stage at high $\sqrt{s}$. The fusion of partons to hadrons or,
inversely, the melting of hadrons to partonic quasiparticles occurs
when the local energy density is about $\varepsilon\approx$
0.5~GeV/fm$^3$. As follows from Fig.~\ref{part}, the parton fraction of
the total energy goes down substantially with decreasing bombarding
energy while the duration of the partonic phase is roughly the same.
The maximal fraction reached is the same in central and peripheral
collisions but the parton evolution time is shorter in peripheral
collisions. One should recall again the important role of the
repulsive mean-field potential for partons in the PHSD model (see
Fig.~\ref{meanF}) that leads to an increase of the flow $v_2$ with
respect to HSD predictions (cf. also Ref.~\cite{CB08}). We point out
in addition that the increase of $v_2$ in PHSD relative to HSD is also
partly due to the higher interaction rates in the partonic medium
because of a lower ratio of $\eta/s$ for partonic degrees of freedom
at energy densities above the critical energy density than for
hadronic media below the critical energy
density~\cite{Mattiello,Bass}. The relative increase in $v_3$ and
$v_4$ in PHSD essentially is due to the higher partonic interaction
rate and thus to a lower ratio $\eta/s$ in the partonic medium, which
is mandatory to convert initial spacial anisotropies to final
anisotropies in momentum space~\cite{Pet4}.

\subsection{Higher-order flow harmonics}

\begin{figure}[b]
\includegraphics[width=\linewidth]{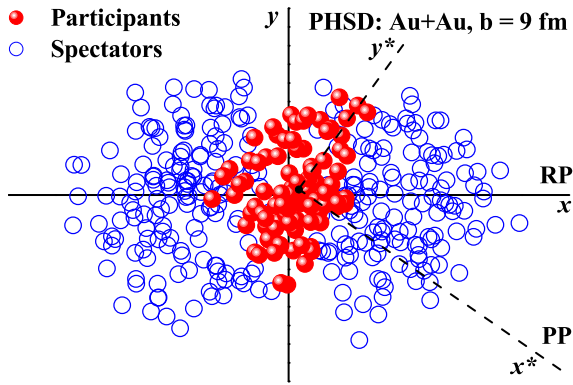}
\caption{(Color online) Projection of a single peripheral Au+Au (200~GeV) collision on
  the transverse plane. Spectator and participant nucleons are plotted
  by empty and filled circles, respectively. The participant plane
  transverse axes are marked by stars ($x^\star,y^\star$).}
\label{C2}
\end{figure}

Depending on the location of the participant nucleons in the nucleus
at the time of the collision, the actual shape of the overlap area may
vary: the orientation and eccentricity of the ellipse defined by the
participants fluctuates from event to event. As seen from
Fig.~\ref{C2}, due to fluctuations the overlap area in a single event
can have, for example, a rotated triangular rather than an almond
shape. Note, however, that by averaging over many events an almond
shape is regained for the same impact parameter.

\begin{figure}[t]
\includegraphics[width=\linewidth]{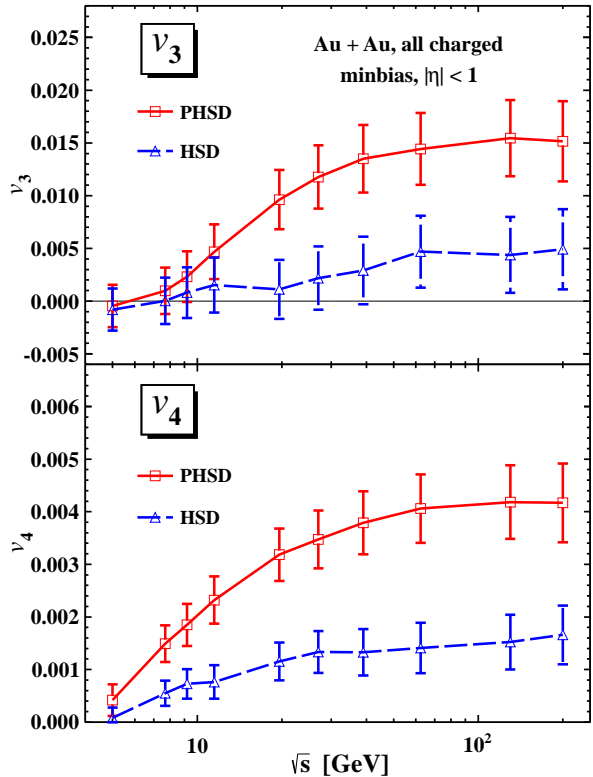}
\caption{(Color online) Average anisotropic flows $v_3$ and $v_4$ of
  charged particles at mid-pseudorapidity for minimum-bias Au + Au
  collisions calculated within the PHSD (solid line) and HSD (dashed
  line) models.}
\label{vns34}
\end{figure}

Recent studies suggest that fluctuations in the initial state geometry
can generate higher-order flow
components~\cite{PHEN07,MMSS08,AR10,Pet123}. The azimuthal momentum
distribution of the emitted particles is commonly expressed in the
form of Fourier series as
\be
&&E\frac{d^3N}{d^3p}= \nonumber \\
&&\frac{d^2N}{2\pi p_Tdp_Tdy}\left(1+\sum^\infty_{n=1} 2v_n(p_T)
\cos [n(\psi-\Psi_n)]\right),\ \ \
\label{eqvn}
\ee
where $v_n$ is the magnitude of the $n$th order harmonic term
relative to the angle of the initial-state spatial plane of symmetry
$\Psi_n$. The anisotropy in the azimuthal angle $\psi$ is usually
characterized by the even-order Fourier coefficients with the reaction
plane $\Psi_n=\Psi_{RP}$: $v_n =\langle exp(\, \imath \,
n(\psi-\Psi_{RP}))\rangle\ ( n = 2, 4, ...)$, since for a smooth
angular profile the odd harmonics vanish. For the odd components, say
$v_3$, one should take into account event-by-event fluctuations with
respect to the participant plane $\Psi_n=\Psi_{PP}$. We calculate the
$v_3$ coefficients with respect to $\Psi_3$ as $v_3\{\Psi_3\} =
\langle \cos(3[\psi-\Psi_3])\rangle/\rm{Res}(\Psi_3)$. The event plane
angle $\Psi_3$ and its resolution $\rm{Res}(\Psi_3)$ are calculated as
described in Ref.~\cite{{AdPH11}} via the two-sub-events
method~\cite{PV98,corrV2}.

In Fig.~\ref{vns34} we display the PHSD and HSD results for the
anisotropic flows $v_3$ and $v_4$ of charged particles at
mid-pseudorapidity for Au+Au collisions as a function of
$\sqrt{s_{NN}}$. The pure hadronic model HSD gives $v_3\approx$ 0 for
all energies. Accordingly, the results from PHSD (dashed red line) are
systematically larger than from HSD (dashed blue line).
Unfortunately, our statistics are not good enough to allow for more
precise conclusions. The hexadecupole flow $v_4$ stays almost constant
in the energy range $\sqrt{s_{NN}}\gsim$ 10~GeV; at the same time the
PHSD gives noticeably higher values than HSD, which we attribute to the
higher interaction rate in the partonic phase, i.e., a lower ratio of
$\eta/s$ for the partonic degrees of freedom~\cite{Mattiello,Bass}.

\begin{figure}[t]
\includegraphics[width=\linewidth]{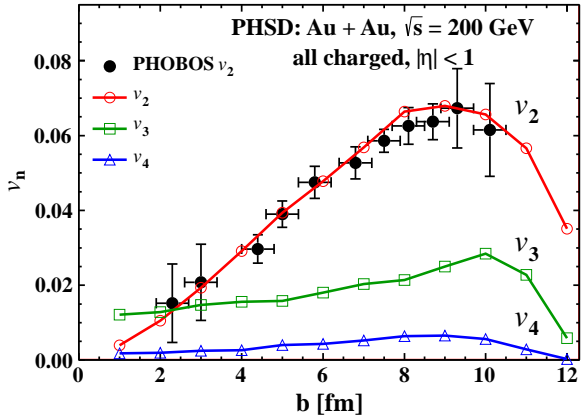}
\caption{(Color online) Impact parameter dependence of anisotropic flows of charged
  particles at mid-pseudorapidity for minimum-bias collisions of Au+Au
  at $\sqrt{s_{NN}}=$ 200~GeV. Experimental points are from
  Ref.~\cite{PHO05}.}
\label{vnb}
\end{figure}

Alongside with the integrated flow coefficients $v_n$ the PHSD model
reasonably describes their distribution over centrality or impact
parameter $b$. A specific comparison at $\sqrt{s_{NN}}=$ 200~GeV
is shown in Fig.~\ref{vnb} for $v_2$, $v_3$, and $v_4$. While $v_2$
increases strongly with $b$ up to peripheral collisions, $v_{3}$ and
$v_{4}$ are only weakly sensitive to the impact parameter. The
triangular flow is always somewhat higher than the hexadecupole
flow in the whole range of impact parameters $b$.

Recently, the triangular flow at $\sqrt{s_{NN}}=$ 200~GeV has been
recalculated in the updated AMPT model~\cite{XK11}. The values of
model parameters for the Lund string fragmentation and the parton
scattering cross section from the previous default version have been
refitted to describe the charged multiplicity distribution, transverse
momentum spectra, and elliptic flow for Au+Au collisions at
$\sqrt{s_{NN}}=$ 200~GeV. In the novel AMPT version the parton
scattering cross sections decrease from about 10 to 1.5 mb. As
compared to the old AMPT result $v_3\approx$0.4, the new value
$v_3\approx$0.2 is consistent with the PHSD results in
Fig.~\ref{vns34}. Note, that the magnitude of the PHSD triangular flow
at $\sqrt{s_{NN}}=$ 200~GeV is similar also to that from the (3 + 1)D
viscous hydrodynamical model~\cite{SJG11} with the specific viscosity
$\eta/s=$0.08. In our calculations the low transverse momentum
particles with $p_T<$ 1~GeV/c are dominating. Unfortunately,
experimental data for this momentum range are not available.

\begin{figure}[t]
\includegraphics[width=\linewidth]{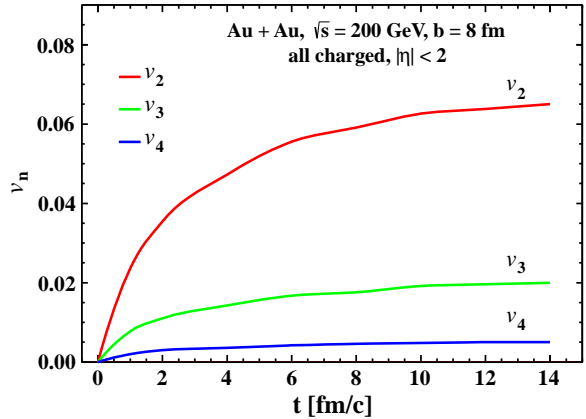}
\caption{(Color online) Time evolution of $v_n$ for Au+Au collisions at
  $\sqrt{s_{NN}}=$ 200~GeV with impact parameter $b =$ 8~fm.}
\label{vntime}
\end{figure}

Figure~\ref{vntime} shows the time evolution of flow coefficients
$v_{2}$, $v_{3}$, and $v_{4}$ for a Au+Au collision at impact
parameter b=8 fm. They reach their asymptotic values by the time of
6-8 fm/c after the beginning of the collision, which corresponds to
the dominantly partonic phase  (cf. Fig.~\ref{part}). Thus,
collective flows are formed in the early partonic stage of the
collision.

\begin{figure}[t]
\includegraphics[width=\linewidth]{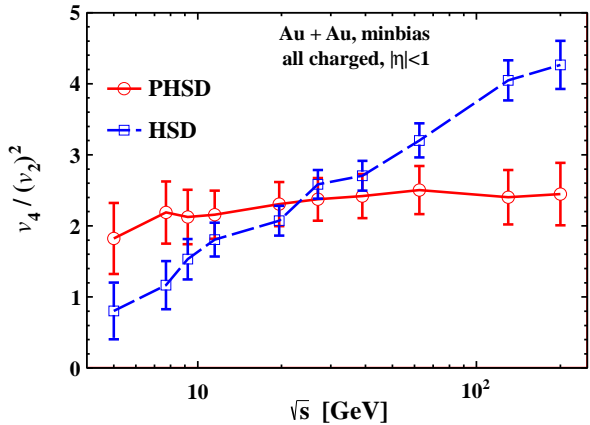}
\caption{(Color online) Beam-energy dependence of the ratio
  $v_4/(v_2)^2$ for Au+Au collisions. The solid and dashed curves are
  calculated within the PHSD and HSD models, respectively.}
\label{v422}
\end{figure}

Different harmonics can be related to each other. In particular,
hydrodynamics predicts that $v_4 \propto (v_2)^2$~\cite{Ko03}. The
simplest prediction that $v_4 = 0.5 (v_2)^2$ is given for a boosted
thermal freeze-out distribution of an ideal fluid, in Ref.~\cite{Bo06}.
In this work it was noted also that $v_4$ is largely generated by an
intrinsic elliptic flow (at least at high $p_T$) rather than the
fourth order moment of the fluid flow. This is a motivation for
studying the ratio $v_4/(v_2)^2$ rather than $v_4$ alone. As is seen
in Fig.~\ref{v422}, indeed the ratio calculated within the PHSD model
is practically constant in the whole range of $\sqrt {s_{NN}}$
considered, but significantly deviates from the ideal-fluid estimate of
0.5. This result is qualitatively consistent with the behavior of
these harmonics in Figs.~\ref{vns} and~\ref{vns34}. In contrast,
neglecting dynamical quark-gluon degrees of freedom in the HSD model,
we obtain a monotonous growth of this ratio.

\begin{figure}[t]
\includegraphics[width=\linewidth]{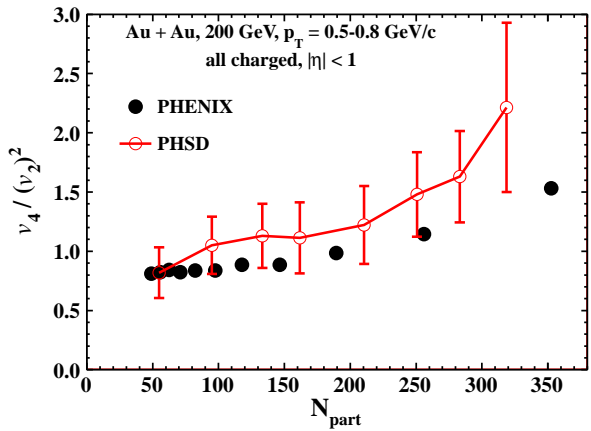}
\caption{(Color online) Participant number dependence of the $v_4/(v_2)^2$ ratio of
  charged particles for Au+Au ($\sqrt{s_{NN}}=$ 200~GeV) collisions.
  The experimental data points for 0.5$<p_T<$0.8~GeV/c are from
  Ref.~\cite{PHENIX_v2_s}.}
\label{v422Np}
\end{figure}

The dependence of the $v_4/(v_2)^2$ ratio versus the number of
participants $N_{part}$ is shown in Fig.~\ref{v422Np} for charged
particles produced in Au + Au collisions at $\sqrt{s_{NN}}=$ 200~GeV.
The PHSD results are roughly in agreement with the experimental data
points from Ref.~\cite{Bai07} but overshoot them for $N_{part}\gsim$
250. We will come back to this quantity in the last subsection when
discussing its $p_T$ dependence.

\subsection{Scaling of flow coefficients}

The $v_2$ coefficient measures the response of the heated and
compressed matter to the spatial deformation in the overlap region
of colliding nuclei, which is usually quantified by the
eccentricity $\epsilon_2=<y^2-x^2>/<x^2+y^2>$. Since the flow response
($v_2$) is proportional to the driving force ($\epsilon_2$), the
ratio $v_2/\epsilon_2$ is used to compare different impact
parameters and nuclei.

\begin{figure}[thb]
\includegraphics[width=\linewidth]{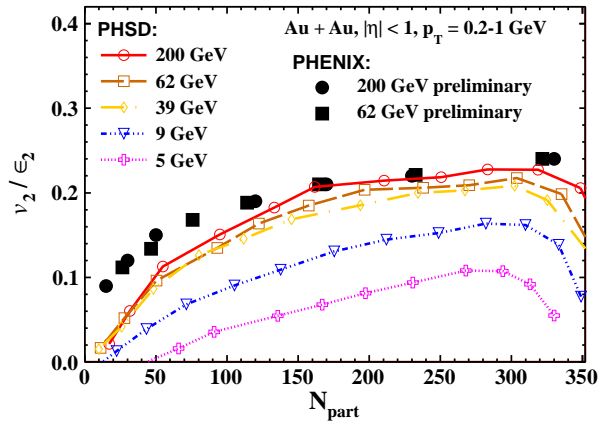}
\caption{(Color online) Scaling of $v_2/\epsilon_2$ as a function of the number of
  participants for different beam energies. Experimental points are
  for Au + Au collisions at $\sqrt{s_{NN}}=$ 200 (circles) and 62~GeV
  (squares)~\cite{Sh11}.}
\label{v2epsB}
\end{figure}

In Fig.~\ref{v2epsB} this ratio is plotted as a function of the
participant multiplicity $N_{part}$. Note that in these calculations
the same eccentricity $\epsilon_2$ was used as in the experiment~\cite{v2v4}.
All PHSD results for $\sqrt{s_{NN}}\gsim$ 40~GeV are very
close to each other and in agreement with experiment. At lower
collision energies this scaling starts to be violated with
decreasing $\sqrt{s_{NN}}$.

\begin{figure}[t]
\includegraphics[width=\linewidth]{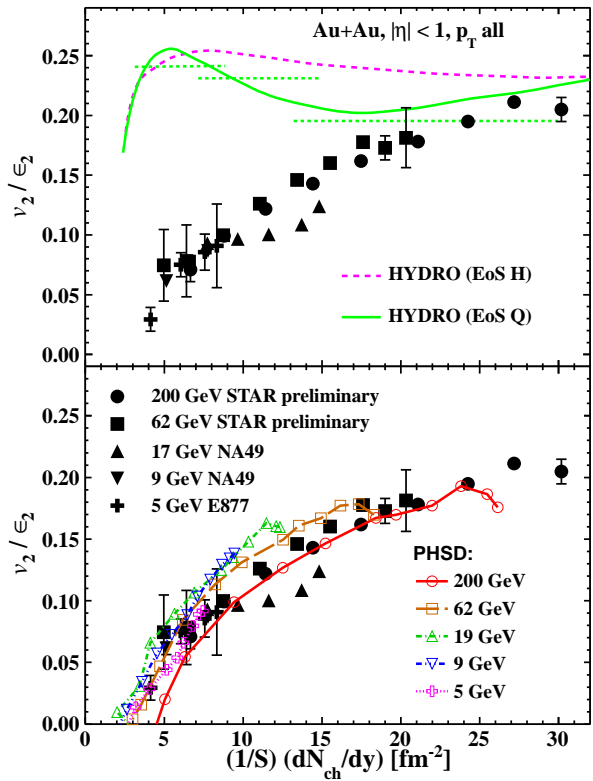}
\caption{(Color online) Scaling of $v_2/\epsilon_2$ vs $(1/S) (dN_{ch}/dy)$. The
  PHSD results are given by lines with open symbols. Predictions of
  ideal boost-invariant hydrodynamics are shown in the top panel
  (from Ref.~\cite{Vol07}) and explained in the text. Our PHSD results are
  presented in the bottom panel. The experimental data points for
  Au+Au collisions at 200~GeV (circles) and 62~GeV (squares) are from
  Refs.~\cite{Vol07,Sh11}.}
\label{v2epsN}
\end{figure}

A remarkable property -- {\it universal scaling} -- has been proposed
in Ref.~\cite{VP00} (see Fig.~\ref{v2epsN}). It appears that
$v_2/\epsilon_2$ plotted versus $(1/S)dN_{ch}/dy$ falls on a
``universal'' curve, which links very different regimes, ranging from
Alternating Gradient Synchrotron (AGS) to RHIC energies. Here
$S=\pi\sqrt{<x^2><y^2>}$ is the overlap area of the collision system
and $dN_{ch}/dy$ is the rapidity density of charged particles.

As seen in Fig.~\ref{v2epsN} very different systems vary essentially
only in a single scale; this scale does not depend on the collision
energy (note that the NA49 data -- filled triangles -- are beyond
this systematics) and is connected to the total entropy
produced~\cite{C06}. Indeed, peripheral and low-energy collisions
are likely to produce systems with incomplete thermalization. Since
rescattering of the particles is rare in the low-density regime,
little change occurs, on average, to the initial momentum
distributions. The measured elliptic flow $v_2$ is therefore
proportional to the initial state eccentricity $\epsilon_2$. This
quantity and the space density of the initial particles $dN_{ch}/dy$
are determined via Glauber calculations. In our case we calculate
these quantities directly within the PHSD model. Thus,
\be
v_2 \propto \frac{1}{S}\frac{dN_{ch}}{dy}\epsilon_2~.
\label{uni}
\ee
We point out that only when event-by-event fluctuations in eccentricity are
taken into account the universal scaling is observed in PHSD. The
term $(1/S)(dN_{ch}/dy)$ contains information about both the longitudinal
structure at freeze-out and the final particle number density (which
is a function of the initial temperature $T$ and baryon chemical
potential $\mu_B$ in the hydro limit).

We therefore use $dN_{ch}/dy$ or, equivalently, the initial entropy
density $s_0$ in central Au+Au collisions as a proxy for the
collision energy: At any given collision energy, a measurement of
$dN_{ch}/dy$ in the most central collision events fixes the value of
$s_0$ to be used in ideal fluid simu\-lations at that energy.
Assuming linear longitudinal expansion without transverse flow at
very early time $\tau_0$, the quantities $dN_{ch}/dy$ and $s_0$ are
thus related by
\be
\frac{dN_{ch}}{dy}\propto \tau_0 s_0~.
\label{entr}
\ee

In an ideal (isentropic) expansion, the final entropy is equal to
the initial entropy content of the system [$\sim$ the initial
particle density $n(T,\mu_B)$]. Thus, the systems from AGS to RHIC
appear to be controlled by a common scale, related to the total
multiplicity, which varies smoothly and drives both $v_2/\epsilon_2$
and $(1/S) (dN/dy)$. This conclusion is a strong indication that
microscopic properties of the system (equation of state and
mean free path) are basically unchanged, up to a shift related to
this scale, in the experimentally addressed energy range.

In the hydrodynamic limit -- implying complete thermalization of the system
-- the ratio of elliptic flow to eccentricity is saturated at very
low impact parameter. In this regime the centrality dependence of
the elliptic flow is mainly determined by the initial elliptic
anisotropy of the overlap zone in the transverse plane, and the ratio
of these two should be approximately constant. This is seen in
Fig.~\ref{v2epsN} where this correlation is plotted by three
horizontal lines for three different beam energies according
to Ref.~\cite{KSH00}.

Predictions of ideal boost-invariant hydrodynamics based on
calculations of Ref.~\cite{KSH00} are also presented in
Fig.~\ref{v2epsN} (top panel). The lines shown are hydrodynamic results
for two boost-invariant lattice-inspired equations of state (with a
quark-hadron phase transition, marked as ``Q'', and for a pure
hadronic system ``H'') calculated for the fixed impact parameter ($b
=$ 7~fm) and different particle densities. Note that hydrodynamic results
do not scale perfectly in this case and in general exhibit a
somewhat flatter centrality dependence at each collision energy. The
deviation from experiment is very large for peripheral collisions
[$(1/S) dN_{ch}/dy\lsim 15$], where the application of hydrodynamics
is questionable since the mean free path of the degrees of freedom
is no longer small compared to the transverse size of the system.

The universal scaling has been investigated in more elaborated
dissipative hydrodynamic models in Refs.~\cite{SH08,Ch11}. A finite
shear viscosity $\eta$ strongly suppresses the buildup of momentum
anisotropy and elliptic flow, especially for low multiplicity
densities. The viscosity effect changes the slope of the
multiplicity scaling for $v_2/\epsilon_2$ but preserves, to a good
approximation, its general scaling with $(1/S)dN_{ch}/dy$. However,
contrary to experiments, in hydrodynamic simulations -- irrespective
of fluid viscosity -- the elliptic flow does not follow the universal
scaling. In principle, the shear viscosity scaled with the entropy
density $\eta/s$ can be extracted in such an analysis; however, the
present accuracy of the extracted values is not high enough.

Thus, the experimentally observed scaling in Fig.~\ref{v2epsN} puts
very strong constraints on the initial microscopic
properties (entropy density, mean free path, {\it etc.}), as well as the
global longitudinal structure~\cite{Tor07}.

\subsection{Scaling in pseudorapidity}

\begin{figure}[t]
\includegraphics[width=\linewidth]{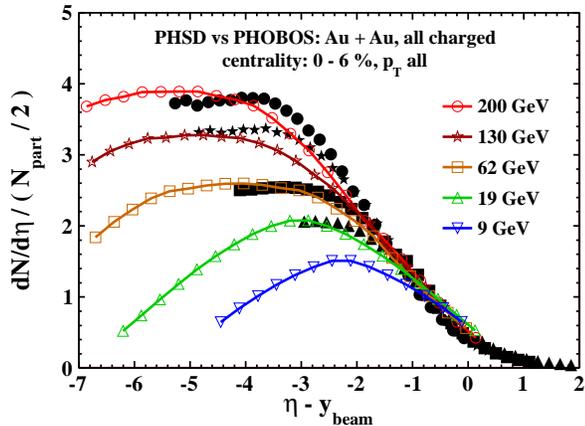}
\caption{(Color online) The dependence of the charged particle multiplicity on the
  shifted pseudorapidity. Experimental data points are from
  Ref.~\cite{Back:2005hs}.}
\label{dNeta}
\end{figure}

Another interesting insight on scale invariance in experimental
observables is the {\it longitudinal scaling} seen experimentally.
The multiplicity longitudinal scaling, $dN_{ch}/d\eta$, is found for
a variety of colliding systems and is often denoted as ``limiting
fragmentation''. Proposed more than 40 years ago by
Feynman~\cite{Fey69} and Hagedorn~\cite{Hag70}, this hypothesis
implies that the multiplicity distribution of particles becomes
independent of $\sqrt{s}$ for $\sqrt{s}\to \infty$. From the
microscopic point of view the multiplicity longitudinal scaling can
be understood if the rapidity distributions of produced hadrons are
functions of the fraction of the hadron longitudinal momentum
$x=2p_T/\sqrt{s}$ alone but not of the total energy. This picture is
very close to the Bjorken scaling of parton distributions. It was
found that models combining ideal hydrodynamics and hadronic
cascades reproduce the longitudinal multiplicity scaling pretty
well, being rather insensitive to the ``phase'' of the system at
thermalization. This is illustrated in Fig.~\ref{dNeta} for very
central Au+Au collisions within the PHSD model. The limiting
fragmentation region is nicely reproduced for $(\eta-y_{beam})\gsim
-2$, while some deviations are seen closer to midrapidity for higher
collision energies. The situation is different for the elliptic
flow, since, unlike $dN_{ch}/d\eta$, the collective flow $v_2$ is
sensitive to the phase of the system as shown before.

\begin{figure}[t]
\includegraphics[width=\linewidth]{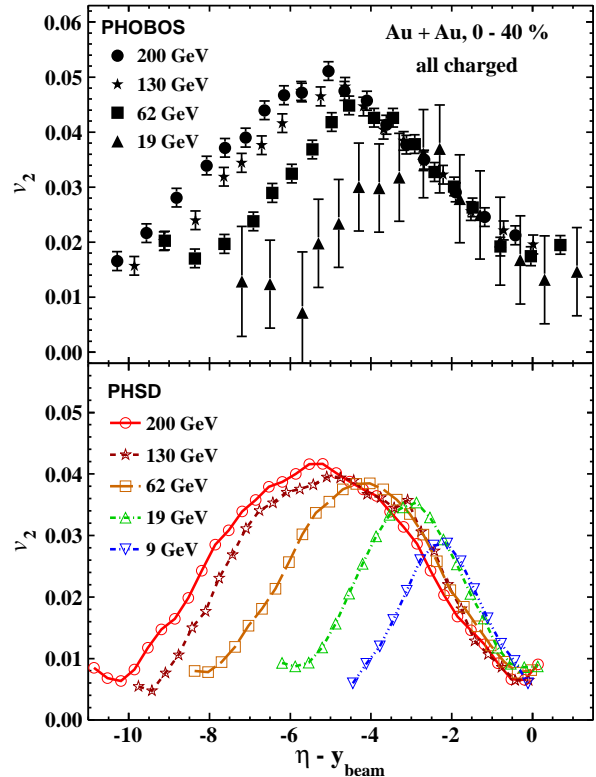}
\caption{(Color online) The $v_2$ dependence versus pseudorapidity as
  calculated in the PHSD model. The compilation of experimental data
  points is from Ref.~\cite{Bu07}.}
\label{etav2e}
\end{figure}

As follows from Fig.~\ref{etav2e} the PHSD model reproduces the
longitudinal $v_2$ scaling up to comparatively low collision
energies (not yet measured). Hadronic results obtained within in the
UrQMD model presented in Ref.~\cite{NJKKNM11} are quite close to our findings, to
be formulated as ``a qualitative agreement with experiment''. Such
a scaling was observed also for partonic models such as the AMPT and
AMPT-SM~\cite{NJKKNM11}.

We mention that the $v_2$ scaling with shifted pseudorapidity can
be obtained also by solving the Boltzmann equation with an
ellipsoidal profile in the initial transverse density~\cite{HL99},
\be
\frac{v_2}{\varepsilon}\sim\frac{<\sigma v>}{S}\
\frac{dN_{ch}}{dy}~,
\label{appv2}
\ee
where $\sigma$ is the interaction cross section.
In line with Eq.~(\ref{uni}) in the discussion of the universal
scaling, such an ansatz will naturally lead to an observed-like
scaling provided that $<\sigma v>$ does not vary with rapidity.

Furthermore, a comprehensive analysis of the longitudinal scaling
has been performed in Ref.~\cite{Tor10} within simple
phenomenological models, trying various assumptions for hydrodynamic
and kinetic descriptions. The authors conclude that the
experimentally observed scaling of {\it multiplicity} with rapidity
and collision energy follows from reasonable models of partonic
dynamics. Neither the free-streaming limit nor the ideal-fluid limit
are expected to break up this multiplicity scaling~\cite{Tor10}. The
situation is, however, different with the scaling observed for the
elliptic flow $v_2$. It is not clear how this scaling could arise
within nonideal hydrodynamics, even if its initial conditions
mirror closely the ones that reproduce the scaling observed in
$dN_{ch}/d\eta$. These remarks address the shape of the scaling
distribution. A more serious problem is the absolute value of $v_2$.
In terms of Eq.~(\ref{appv2}), to get a reasonable magnitude of
$v_2$, the cross section $\sigma$ should be increased to the point
where the Knudsen number is well below unity~\cite{Tor10}. As
demonstrated above, the PHSD model allows us to get reasonable results
for the multiplicity and $v_2$ longitudinal scaling by default
without any tuning of parameters.

\subsection{Differential distributions}

It was shown in the RHIC experiments that, for a given centrality, the
differential elliptic flow for all observed hadrons scales to a single
curve when plotted as $v_2/n_q$ versus $E_{TK}/n_q$, where $n_q$ is the
number of constituent quarks in a given hadron species and $E_{TK}$ is
the transverse kinetic energy for these
hadrons~\cite{LT06,PHENIX07}. This quark number scaling is consistent
with the recombination model, which assumes the collective flow to
develop on the quark level in the QGP phase. Such a scaling of the
elliptic flow $v_2$ for identified hadrons has been measured by the
STAR and PHENIX Collaborations at different centralities for top RHIC
energies and recently also within the BES program~\cite{ST11}. Our
paper deals only with charged-particle observables and therefore the
number of constituent quarks $n_q$ is not accurately defined. The
analysis of scaling properties of identified hadrons we postpone to a
future study.

\begin{figure}[t]
\includegraphics[width=\linewidth]{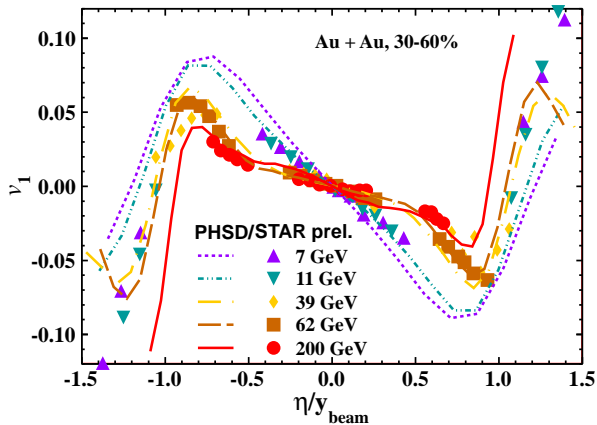}
\caption{(Color online) Normalized pseudo-rapidity distributions of
  the directed flow $v_1$ in the transient collision energy range. The
  experimental data points are from the STAR
  Collaboration~\cite{Pa11}.}
\label{v1d}
\end{figure}

We start with the consideration of rapidity distributions
$dv_1/d\eta$ of the directed flow~\cite{Pa11} and its beam energy
dependence as presented in Fig.~\ref{v1d}. The directed flow $v_1$
is the first harmonic coefficient of the above Fourier expansion of
the final momentum-space azimuthal anisotropy Eq.~(\ref{eqvn}), and
it reflects the collective sidewards motion or ``bounce-off'' of the
particles in the final state. Being generated essentially during the
nuclear passage time $\sim 2R/\gamma $, the directed flow probes
 the very  early stage of the collision  dynamics. In the region
closer to the beam/target rapidity than to midrapidity, the directed
flow is generated very early even at a pre-equilibrium stage of the
collision~\cite{So97} and thus it probes the onset of bulk
collective behavior. In Fig.~\ref{v1d} the directed flow of charged
particles is plotted versus the normalized pseudorapidity
$\eta/y_{beam}$ in the large range of the BES collision energies for
centrality 30\%-60\%. We observe that $v_1(\eta/y_{beam})$ shows a
beam-energy scaling behavior, though not perfect. Both hydrodynamic
and nuclear transport models indicate that the directed flow is a
sensitive signature for a possible phase transition, especially in
the central region of beam-energies under investigation. In
particular, the shape of $v_1(y)$ in the midrapidity region is of
special interest because it has been argued that differential
directed flow may exhibit flatness at midrapidity due to a strong,
tilted expansion of the source. Such tilted expansion gives rise to
antiflow~\cite{CR99}. The antiflow is in the opposite direction
to the repulsive bounce-off motion of nucleons. If the tilted
expansion is strong enough, it can cancel and even reverse the
motion in the bounce-off direction and result in a negative $v_1(y)$
slope at midrapidity, potentially producing a wiggle-like structure
in $v_1(y)$. A wiggle for baryons is a possible signature of a phase
transition between hadronic matter and quark gluon plasma (QGP),
although a QGP is not the only possible
explanation~\cite{CR99,Br00,St05}. As seen from Fig.~\ref{v1d} the
slope of the $v_1(\eta/y_{beam})$ distribution at $\eta=0$ is
negative and stays almost constant for $\sqrt{s_{NN}}\gsim$ 10~GeV;
its magnitude slightly increases with decreasing beam enery,
however, exhibiting no irregularities.

\begin{figure}[t]
\includegraphics[width=\linewidth]{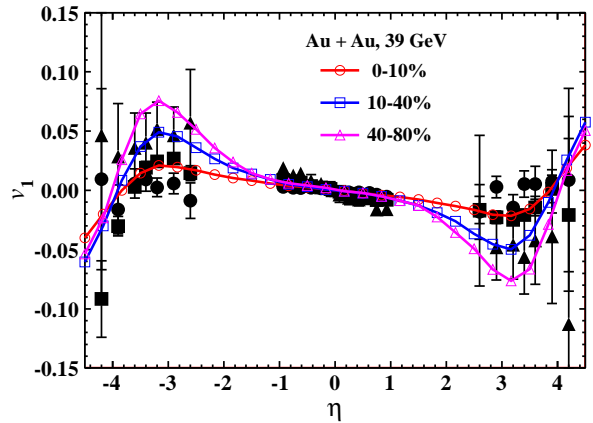}
\caption{(Color online) Directed flow distributions at different centrality for Au +
  Au collisions at $\sqrt{s_{NN}}=$ 39~GeV. The experimental data
  points are from Refs.~\cite{Pa11,Abelev:2008jga}.}
\label{v1centr}
\end{figure}

The slope of the pseudorapidity distributions is slightly changed
when different criteria for centrality selection are applied as
demonstrated in Fig.~\ref{v1centr}. The influence of this selection
is very moderate at midrapidity but becomes noticeably stronger in
the target-projectile fragmentation region with increasing impact
parameter.

\begin{figure}[thb]
\includegraphics[width=\linewidth]{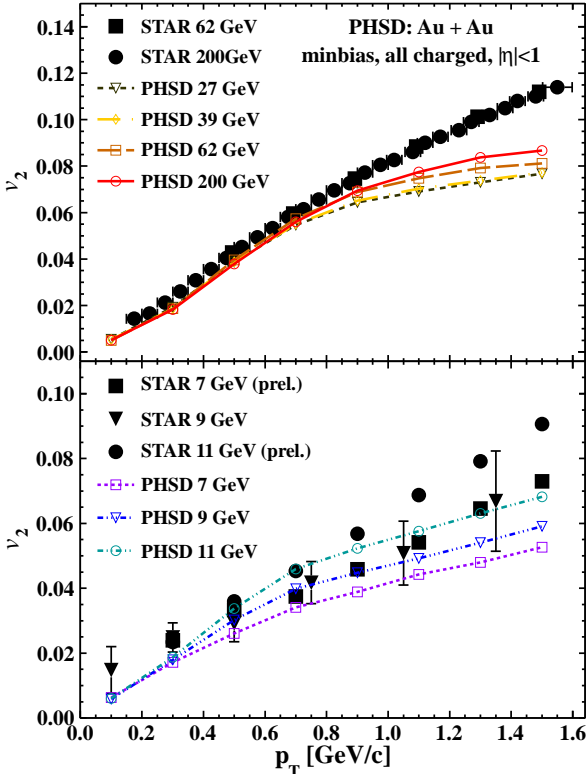}
\caption{(Color online) Beam-energy evolution of transverse momentum
  distributions of $v_2(p_T)$ for Au+Au collisions in comparison to
  the data of the STAR Collaboration from
  Refs.~\cite{AbST10,STARv2pt,Pa11}.}
\label{Pt}
\end{figure}

Let us continue with differential distributions of the elliptic flow
$v_2$ by comparing the $p_T$ dependence from data with those from the
PHSD model. The results from PHSD for $v_2(p_T)$ are displayed in
Fig.~\ref{Pt} for $\sqrt{s_{NN}}$ from 5 to 200~GeV. Also shown are
the corresponding results from the STAR Collaboration at
$\sqrt{s_{NN}}=$ 9, 62, and 200~GeV (by symbols). The data from PHENIX
and STAR at midrapidity indicate that the magnitudes and trends of the
differential elliptic flow [$v_2( p_T)$, centrality dependence], are
changed only very little over the collision energy range
$\sqrt{s_{NN}}=$ 62 - 200~GeV, indicating an approximate saturation
of the excitation function for $v_2$ at these energies~\cite{Tar11} as
exemplified in Fig.~\ref{Pt}. We mention that the PHSD results
underestimate the data systematically for $p_T >$ 1~GeV, which is
attributed to an overestimation of scattering of partons with high
transverse momenta. However, the collective flow $v_2$ of the ``bulk
matter'' is rather well described at all energies without any tuning of
parameters.

\begin{figure}[t]
\includegraphics[width=\linewidth]{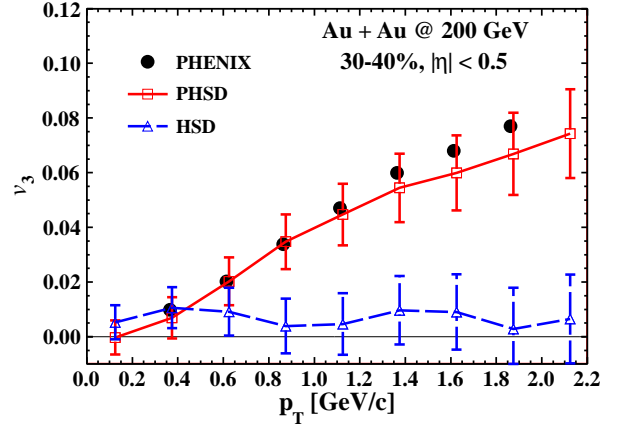}
\caption{(Color online) Triangular flow $v_3$ as a function of transverse momentum
  $p_T$ for Au + Au collision at $\sqrt{s_{NN}}=$ 200~GeV. The data
  points are from Ref.~\cite{AdPH11}.}
\label{v3Pt}
\end{figure}

The momentum distribution of the flow $v_3$ is also in a quite
reasonable agreement with experiment (see Fig.~\ref{v3Pt}).

\begin{figure}[t]
\includegraphics[width=\linewidth]{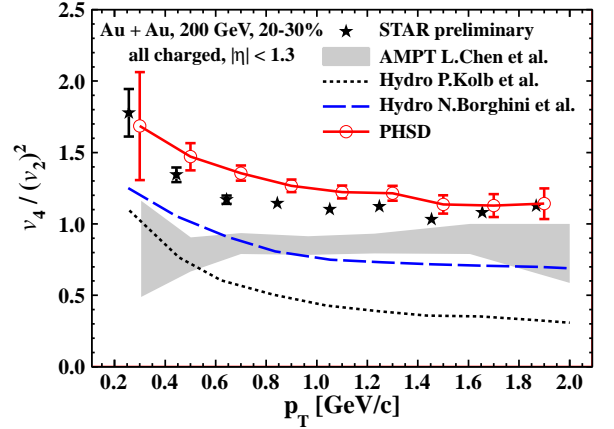}
\caption{(Color online) Transverse momentum dependence of the ratio $v_4/(v_2)^2$ of
  charged particles for Au+Au (at $\sqrt{s_{NN}}=$ 200~GeV)
  collisions. The dashed and dot-dashed lines are calculated within
  the hydrodynamic approaches from Refs.~\cite{Ko03} and~\cite{BO06},
  respectively. The shaded region corresponds to the results from the
  AMPT model~\cite{CKL04}. The experimental data points are from the
  STAR Collaboration~\cite{Bai07}.}
\label{v422pt}
\end{figure}

As pointed out before, the ratios of flow coefficients might shed
valuable light on the actual dynamics, since especially the ratio
$v_4/(v_2)^2$ is sensitive to the microscopic dynamics. In this
respect we show the transverse momentum dependence of the ratio
$v_4/(v_2)^2$ in Fig.~\ref{v422pt} for charged particles produced in
Au+Au collisions at $\sqrt{s_{NN}}=$ 200~GeV (20-30\% centrality).
The PHSD results are quite close to the experimental data points from
Ref.~\cite{Bai07}; however, they overestimate the measurements by up
to 20\%. The hydrodynamic results -- plotted in the same figure --
significantly underestimate the experimental data and noticeably
depend on viscosity. The partonic AMPT model~\cite{CKL04} discussed
above also predicts a slightly lower ratio than the measured one;
however, it is in agreement with both hydrodynamic models for
$p_T\lsim$ 0.8~GeV/c. Our interpretation of Fig.~\ref{v422pt} is as
follows: the data are not compatible with ideal hydrodynamics and a
finite shear viscosity is mandatory (in viscous hydrodynamics) to come
closer to the experimental observations. The kinetic approaches AMPT and
PHSD perform better but either overestimate (in AMPT) or slightly
underestimate (in PHSD) the scattering rate of soft particles. An
explicit study of the centrality dependence of these ratios should
provide further valuable information.

\section{Conclusions}
\label{sec4}

In summary, relativistic collisions of Au+Au from $\sqrt{s_{NN}}=$ 5
to 200~GeV have been studied within the PHSD approach, which
includes the dynamics of explicit partonic degrees of freedom as
well as dynamical local transition rates from partons to hadrons and
also the final hadronic scatterings. Whereas earlier studies have
been carried out for longitudinal rapidity distributions of various
hadrons, their transverse mass spectra, and the elliptic flow $v_2$
as compared to available data at SPS and RHIC
energies~\cite{PHSD,BCKL11}, here we have focused on the PHSD
results for the collective flow coefficients $v_1$, $v_2$, $v_3$, and
$v_4$ in comparison to recent experimental data in the large energy
range from the RHIC Beam Energy Scan (BES) program, as well as
different theoretical approaches ranging from hadronic transport
models to ideal and viscous hydrodynamics. We mention explicitly
that the PHSD model from Ref.~\cite{BCKL11} has been used for all
calculations performed in this study and no tuning (or change) of
model parameters has been performed.

We have found that the anisotropic flows -- elliptic $v_2$, triangular
$v_3$, and hexadecapole $v_4$ -- are reasonably described within the PHSD
model in the whole transient energy range, naturally connecting the
hadronic processes at lower energies with ultrarelativistic collisions
where the quark-gluon degrees of freedom become dominant. The smooth
growth of the elliptic flow $v_2$ with the collision energy
demonstrates the increasing importance of partonic
degrees of freedom. This feature is reproduced by neither
hadron-string based kinetic models nor a multi phase transport (AMPT)
model treating the partonic phase in a simplified manner. Other
signatures of the transverse collective flow, the higher-order
harmonics of the transverse anisotropy $v_3$ and $v_4$, change only
weakly from $\sqrt{s_{NN}}\sim$ 7~GeV to the top RHIC energy of
$\sqrt{s_{NN}}=$ 200~GeV, roughly in agreement with experiment. As
shown in this study, this success is related to a consistent treatment
of the interacting partonic phase in PHSD, whose fraction increases
with the collision energy.

The observables calculated within the PHSD model exhibit some scaling
properties for collision energies above $\sqrt{s_{NN}}=$ 40~GeV. In
particular, the universal scaling of $v_2/\epsilon_2$ versus $(1/S)
dN_{ch}/dy$ ({\it cf.} Fig.~\ref{v2epsN}) is approximately reproduced
as well as the longitudinal scaling of the charged particle
pseudorapidity distributions of the elliptic flow $v_2$ in $(\eta -
y_{beam})$ representation ({\it cf.} Figs.~\ref{dNeta} and
\ref{etav2e}) in this energy range. This feature is not reproduced by
hadronic transport models (such as HSD and UrQMD) and meets (severe)
problems in the various hydrodynamic descriptions.

The analysis of correlations between particles emitted in
ultrarelativistic heavy-ion collisions at large relative rapidity has
revealed an azimuthal structure that can be interpreted as being
solely due to collective flow~\cite{TY10,LGO10}. This interesting new
phenomenon, denoted as triangular flow, results from initial-state
fluctuations and a subsequent hydrodynamic-like evolution.  Unlike the
usual directed flow, this phenomenon has no correlation with the
reaction plane and should depend weakly on rapidity.  Event-by-event
hydrodynamics~\cite{GGH11} has been a natural framework for studying
this triangular collective flow but it has been of interest also to
investigate these correlations in terms of the PHSD model. We have
found the third harmonics to increase steadily in PHSD with bombarding
energy. The coefficient $v_3$ is compatible with zero for
$\sqrt{s_{NN}} >$ 20~GeV in case of the hadronic transport model HSD,
which does not develop ``ridge-like'' correlations. In this energy range
PHSD gives a positive $v_3$ due to dominant partonic interactions.

Different harmonics can be related to each other and, in particular,
hydrodynamics predicts that $v_4 \propto (v_2)^2$~\cite{Ko03}. In
this work it was noted also that $v_4$ is largely generated by an
intrinsic elliptic flow (at least at high $p_T$) rather than the
fourth-order moment of the fluid flow. Indeed, the ratio
$v_4/(v_2)^2$ calculated within the PHSD model is approximately
constant in the whole considered range of $\sqrt {s_{NN}}$ but
significantly deviates from the ideal fluid estimate of 0.5. In
contrast, neglecting dynamical quark-gluon degrees of freedom in the
HSD model, we obtain a monotonous growth of this ratio.

The transverse momentum dependence of the ratio $v_4/(v_2)^2$ at the
top RHIC energy has given further interesting information ({\it cf.}
Fig.~\ref{v422pt}) by comparing the various model results to the data
from STAR, which are interpreted as follows: the STAR data are not
compatible with ideal hydrodynamics and a finite shear viscosity is
mandatory (in viscous hydrodynamics) to come closer ot the
experimental ratio observed. The kinetic approaches AMPT and PHSD
perform better but either overestimate (in AMPT) or slightly
underestimate the scattering rate of soft particles (in PHSD). An
explicit study of the centrality dependence of these ratios should
provide further valuable information.

It will be promising to extend our studies to asymmetric heavy-ion
collisions that can be used to constrain models dealing with flow
fluctuations in heavy-ion collisions but with a larger sensitivity
for $v_2$-related observables than for $v_3$~\cite{HNM11}.
Independently, an extension of the PHSD approach to LHC
energies with possible color-glass-condensate initial conditions
has to be performed in future.

\section*{Acknowledgements}

We are thankful to O.~Linnyk and G.~Torrieri for helpful
discussions. This work was supported in part by the DFG Grants No.\ WA
431/8-1 and No.\ CA 124/7-1, the RFFI Grants No.\ 08-02-01003-a, the
Ukrainian-RFFI Grant No.\ 09-02-90423-ukr-f-a, and the LOEWE center HIC for
FAIR.

\end{document}